# A quantum pathway to overcome the trilemma of magnetic data storage

Patrick Robert Forrester[1,2,#], François Patthey[1], Edgar Fernandes[1], Dante Philippe Sblendorio[1], Harald Brune[1*], and Fabian Donat Natterer[1,2*]

1 Institute of Physics, École Polytechnique Fédérale de Lausanne, CH-1015 Lausanne, Switzerland
2 Physik-Institut, University of Zurich, CH-8057 Zurich, Switzerland
# Present address: Physical Measurement Laboratory, National Institute of Standards and Technology, Gaithersburg, MD 20899, USA; Institute for Research in Electronics and Applied Physics & Maryland NanoCenter, University of Maryland, College Park, MD 20742, USA
*To whom correspondence should be addressed: fabian.natterer@uzh.ch and harald.brune@epfl.ch

**The three essential pillars of magnetic data storage devices are readability, writeability, and stability. However, these requirements compete as magnetic domain sizes reach the fundamental limit of single atoms** [1–5] **and molecules** [6–9]**. The proven magnetic bistability of individual holmium atoms on magnesium oxide** [1,2,10] **appeared to operate within this magnetic trilemma, sacrificing writeability for unprecedented stability** [10]**. Using the magnetic stray field created by the tip of a spin-polarized scanning tunneling microscope (SP-STM), we controllably move the Ho state into the quantum regime, allowing us to write its state via the quantum tunneling of magnetization (QTM)** [11]**. We find that the hyperfine interaction causes both the excellent magnetic bistability, even at zero applied magnetic field, and the avoided level crossings which we use to control the magnetic state via QTM** [12]**. We explore how to use such a system to realize a high-fidelity single atom NOT gate (inverter). Our approach reveals the prospect of combining the best traits of the classical and quantum worlds for next generation data storage.**

Big data has shown early promise towards solving diverse problems ranging from medicine and basic science to alternative energy and transportation [13,14]. When coupled with modern machine learning algorithms, otherwise intractable problems are solved while minimizing energy consumption and human labor [13]. This approach, however, demands high-fidelity and high density data storage devices [14]. The traditional model of scaling down classical memory components by reducing magnetic domain sizes faces roadblocks due to writeability limitations [15]. It was recently proven that this trend to smaller bit sizes could be continued to the fundamental limit of individual molecules [6–9], atomic clusters [3,4] and single atoms [1,2,5,16]. However, this continuation comes at a cost: the magnetic trilemma that creates competition between requirements for reading, writing, and retaining data appeared insurmountable in atomic scale systems [15]. This is owed to the large magnetic anisotropy energy (MAE) barrier [17] that protects bits against thermal state reversal. Through the clever consideration of symmetry arguments [18,19], single atomic and molecular systems with large MAEs and suppressed quantum tunneling of magnetization (QTM) have been discovered [1,8–10,12,16]. Unfortunately, the magnetic switching field required to write bits doubles with every factor of two reduction in bit size. Indeed, the coercive field of Ho single atom magnets exceeds 8 T [10] far beyond technically accessible field magnitudes of common hard disc drive (HDD) write heads. Roadmaps for future HDD technologies propose thermally [20] and microwave [21] assisted magnetic recording that briefly lower the coercive field. However, these approaches are limited to domain densities smaller than 20 Terabit/in$^2$ [15] far from the already demonstrated 115 Terabit/in$^2$ density of arrays of single atom magnets [16]. We therefore introduce a different paradigm combining the stability of large MAEs with the utility of controlled QTM for data control, demonstrating the feasibility of practical data storage at the fundamental limit.

We show our approach on the stable single atom magnet holmium on magnesium oxide [1,2,10,22]. In this study, we first demonstrate the excellent magnetic bistability of Ho single atom magnets by spin-polarized scanning tunneling microscopy (SP-STM) [23] (Figure 1). The two total spin ground state orientations of the Ho moment exhibit tunneling magnetoresistance (TMR) of several percent when measured with SP-STM [2,10,22]. As was shown previously, the magnetic states of Ho are bistable when measured at threshold voltages below $V_t$=73 mV in an applied out-of-plane magnetic field [2,10,22]. The magnetic state becomes unstable above this threshold and switches randomly between its two magnetic orientations (denoted Up and Down), as exemplified by the current-time traces in Figure 1(b, $V_2$>$V_t$). The switching traces are labelled by 'read' to emphasize their utility in determining the magnetic state of the Ho atom before varying external parameters, such as bias voltage and magnetic field. In Figure 1, we leave Ho in the Up state (left), then drop the bias voltage below the switching threshold for the 'evolve' phase ($V_1$<$V_t$, middle), and observe the Ho state above the switching threshold again in the Up state in the final 'read' phase (right, $V_2$>$V_t$) some minutes later. These measurements have been repeated numerous times to support the observation of excellent stability of the Ho state below the switching threshold. Unlike previous reports [2,10], however, our measurements are carried out without an applied external magnetic field, relying on the intrinsic SP of an antiferromagnetic STM tip [22]. Although our measurements suggest magnetic bistability at zero field, the tip may still produce a small, nonzero magnetic stray field, $B_0$, [24,25], and it is unclear from these initial measurements whether the tip field is responsible for the system's stability.

To more rigorously test for zero-field stability, we employ a measurement protocol that reduces the tip-stray field to effectively zero during the 'evolve' phase and allows us to read the state of the Ho atom after evolution under tunneling conditions (with the stray field of the tip possibly stabilizing the magnetic state). As above, we first identify the Ho state in the 'read' phase. We then retract our tip sufficiently far that the tip stray field experienced by the Ho atom is vanishingly small (less than the earth's magnetic field), allowing the Ho atom to evolve in the absence of a magnetic field. We then place the tip back onto the Ho atom to check its state again.

Figure 2 summarizes our experimental protocol. To identify the initial state, we begin with the tip on the Ho atom at a bias voltage above the switching threshold (purple 'read' area, $V_2>V_t$, $B=B_0$). The bias voltage is then lowered below the switching threshold ($V_1<V_t$), and the tip is pulled back (yellow 'evolve' area, $B=0$). We keep the tip removed for some time, $\Delta t$, during which the atom is free to evolve. We next return the tip back onto the Ho atom while keeping the bias voltage below the switching threshold ($V_1<V_t$, $B=B_0$). To measure the Ho state after the free evolution, we increase the bias voltage above the switching threshold (green 'read' area, $V_2>V_t$, $B=B_0$).

There are four possible outcomes for such an experiment (read, evolve, read): The Ho atom either retains its initial state (Up stays Up or Down stays Down), switches from Up to Down, or switches from Down to Up (Fig. 2b). We repeat this experiment (read, evolve, read) for various $\Delta t$ values 884 times with 5 different tips and plot the respective occurrence of switches in Figure 2c (left). Most notable is the overwhelming observation of state conservation. We see that the Ho atoms retain their initial state, irrespective of whether they started in the Up ($99^{+1}_{-2}\%$) or Down [$(91 \pm 5)\%$] state. Further, this zero-field stability is independent of $\Delta t$, the time during which the Ho atom is free to evolve. This initial experiment proves the excellent magnetic bistability of Ho atoms at zero field. However, we do observe a statistically significant occurrence of state reversals, which must have a different origin.

To investigate whether our magnetic field sweeps, carried out by moving the tip from tunneling distances to full retraction, $\Delta Z$, induce the observed state reversals, we modify the experiment sketched in Figure 2a. This time, we repeatedly retract and approach the tip in the 'evolve' phase before reading the state (dotted blue trace in Figure 2a upper, $V_1<V_t$), instead of merely holding the tip far away from the Ho atom. This is equivalent to several magnetic field sweeps from $B=B_0$ to $B=0$ and back, since the tip exposes the Ho atom to some finite stray field, $B_0$, at tunneling conditions. We plot the occurrence of switches in Figure 2d. A single retract/approach cycle (*N*=1) results in the same negligible occurrence of state reversals: $0^{+3}_{-0}$% (starting Up) and $3^{+6}_{-3}$% (starting Down). However, when we increase the number of retract/approach cycles in the 'evolve' phase (*N*=20), we see the occurrence of magnetic state reversals grow to (2±1)% starting Up and (30±3)% starting Down (Figure 2c). For even more retract/approach cycles (*N*=100), we observe a similar occurrence: $3^{+2}_{-1}$% (starting Up) and $28^{+5}_{-4}$% (starting Down). This experiment proves that the magnetic state reversals are caused by the sweeping of the magnetic field.

Earlier models constrained the magnetic ground state of Ho on MgO [1,10] to two possible out-of-plane projections of the total angular momentum, namely $J_z$=8 and $J_z$=7 [10], but it remained unclear whether the magnetic bistability originated from crystal field symmetry or an applied magnetic field. These models also neglected the hyperfine interaction, which we now explore using the EasySpin toolbox [26]. We compare the behavior of these two suggested ground state models to our observations, including the hyperfine interaction for a nuclear spin of *I* = 7/2 (100% natural abundance). We use the hyperfine coupling constant (*A* ≈ *900* MHz) previously measured for similar systems [12,27] and the reported Steven's parameters of Ref [10]. Figure 3(a,d) show the resulting extended Zeeman diagrams for the two possible $J_z$=8 and $J_z$=7 models. The insets in panels (a,d) describe the zero-field level diagrams with the respective lowest energy electronic ground states indicated by larger dots [10]. When we include the nuclear spin of Ho, the total spin states are each split into (2*I*+1)=8 non-degenerate states.

At zero field, states with opposite nuclear and total electron spin meet, resulting in 8 zero field level crossings. Figure 3b (solid orange) shows a representative zero field crossing for the $J_z=8$ model. The hyperfine interaction leads to a real level crossing at zero field (solid orange in Figure 3b), while its absence would have resulted in an avoided level crossing and substantial state mixing (dashed red in Figure 3b), as was suggested previously [10]. Figure 3e shows the same real zero-field crossing for the $J_z=7$ model. We observe similar characteristics for all zero-field crossings, preliminarily explaining the observed magnetic bistability. However, since this bistability would occur for both total spin models, we investigate below how to ultimately identify the electronic ground state.

When examining the $J_z=8$ model, we notice several avoided level crossings at nonzero field values. Figure 3c shows one such avoided level crossing ($B_1$) at which the electronic states exhibit significant mixing and a tunnel splitting, $\Delta$, of 0.1 µeV. We observe 8 such crossings, all with a similar splitting (indicated by red circles in Figure 3a). The $J_z=7$ model, on the other hand, is devoid of any avoided level crossings. We show a representative real crossing for the $J_Z=7$ model in Figure 3f for reference. Sweeping the magnetic field through an avoided level crossing results in a nonzero probability of magnetic state reversal [11]. During a magnetic field sweep, a Ho state in the $J_z=7$ case would never encounter an avoided level crossing. But we do observe an increased occurrence of magnetization reversals when we repeatedly sweep our magnetic field (retract/approach tip). Therefore, we conclude that the Ho ground state must be $J_z=8$.

To confirm the existence of avoided level crossings and to demonstrate their potential utility, we repeat the experiment described in Figure 2a. This time, we retract and approach the tip such that the stray field varies linearly in time ($dB/dt$ = constant) using an extremely fast (~$10^3\ T/s$) magnetic field sweep rate when moving the tip in one direction (retracting or approaching the tip) and a much smaller (~ $5 \times 10^{-3}\ T/s$) sweep rate when moving the tip in the opposite direction. If state reversal was caused by QTM at an avoided level crossing, this protocol should result in a magnetic state inversion, as the probability of magnetic state reversal according to Landau-Zener [28] is proportional to $p = 1 - \exp(-\alpha\Delta^2/\frac{dB}{dt})$, where α is some constant. This means that sweeping adiabatically will flip the state (total spin) and sweeping quickly will retain the state. Performing a fast and slow sweep consecutively should result in a net reversal. Table 1 shows the incidence of observed magnetic state reversal for these protocols. Both show an occurrence of magnetic state reversal on the order of 50% regardless of the initial state, as predicted by the Landau-Zener theory. Note, we do not expect occurrences near 100% as a maximum of 4 out of the 8 crossings are traversed when sweeping the magnetic field from some positive value, $B_0$, to zero. We observe slightly different occurrences of magnetic state reversal depending on the initial Ho state, as expected given the difference in the thermal occupation distributions for the Up and Down spin manifolds. This explanation is corroborated by the efficiency of magnetic state reversal for each initial Ho state reversing when the sweep order is reversed. In other words, Up to Down for the slow-fast protocol maps to Down to Up for the fast-slow protocol. If the field sweep was extended from $B_0$ to $-B_0$, this would facilitate a high-fidelity state inverter that could serve as a general scheme to control the magnetic state of single ion magnets.

We have established a proof of principle experiment that clearly demonstrates our ability to deliberately invert and thus write the magnetic state of a single atom magnet via the quantum tunneling of magnetization at an avoided level crossing. By using the excellent magnetic bistability of Ho at virtually all magnetic field values and the convenient writeability near avoided level crossings via QTM, we effectively overcome the magnetic trilemma of data storage for the smallest realizable bit size. Our approach removes the need for high energy tunneling electrons, since the magnetization reversal can be fully induced by a controlled state mixing at an avoided level crossing. This technique should be portable to insulating substrates resulting in superior isolation from the scattering with itinerant electrons believed to be limiting other single atom and molecule magnets [1,2,29]. The fast magnetic field sweep rates used here are easily accessed by conventional HDD write heads [30]. We anticipate the adoption of using large MAE systems for data conservation and QTM for data manipulation to facilitate single molecule and atom magnetic data storage at ultimate data densities. Avoided level crossings furthermore represent an ideal toy system for exploring conventional and quantum logic, which we touch on in this STM study. The former would realize the fundamental size limit of a logic gate, and the latter would provide invaluable insight in developing more robust qubits.

**Methods**

**STM measurements.** All measurements are performed using a homemade low-temperature STM, operating at a pressure of $1 \times 10^{-10}$ mbar and temperature of 4.7 K [31]. Following previous studies [1,10,22,25], we grow ~1.5 monolayers of MgO by exposing an atomically clean Ag(100) crystal, held at ~773 K, to an Mg flux from a Knudsen cell evaporator in an oxygen partial pressure of ~$1.33 \times 10^{-6}$ mbar at a growth rate of ~0.2 monolayers per minute. We dose Ho atoms directly onto the cooled sample (~10 K) using a thoroughly degassed *e*-beam evaporator. All STM measurements are taken with an antiferromagnetic $Mn_{88}Ni_{12}$ tip prepared using the recipe in reference [22]. Spin-polarization is verified via voltage dependent switching analysis of Ho [2,10,22]. The STM tip is retracted to distances of 5 nm and 1 nm for the single retract/approach and multiple retract/approach experiments, respectively. For all $\Delta t$ values except 60 s, the tip was retracted in one step, limited by the slew rate of control electronics and the z-piezo (~microseconds). For the 60 s data and the multiple retract/approach cycles, the tip is retracted linearly in 100 ms. For the state inversion measurements, the tip is retracted 1.3 nm over the course of 20 s such that $dB/dt$ is constant. The single retract/approach experiment was performed 884 times with 5 different tips, the multiple retract/approach cycles 691 times with four different tips, and state inversion experiment repeated 110 times with the same tip.

**State reversal analysis.** We correct for unseen switching events that arise from limited preamplifier bandwidth in Figure 2c. We assume a Markovian probability of state reversal and calculate the expectation value of state reversal during the time that falls outside of the temporal resolution of our experiment using the measured residence times of the Up and Down states. We subtract these reversals from the total measured state reversals. Error bars represent confidence intervals of one standard deviation calculated using the Agresti-Coull method for binomial processes.

**Spin Hamiltonian Calculations.** To construct the extended Zeeman diagrams and total spin level diagrams for the two total spin models, we employ the MATLAB toolbox EasySpin V5.2.23 [26] to implement the spin Hamiltonian, $H = H_{cf} + H_{HF} + H_z$, where $H_{cf} = B_0^2 O_0^2 + B_0^4 O_0^4 + B_4^4 O_4^4 + B_0^6 O_0^6 + B_4^6 O_4^6$, $H_{HF} = A_{HF} JI$, and $H_z = -g_{Eff} J_Z B_Z \mu_B$. Here, $\{O_m^n\}$ are the crystal field operators allowed by the fourfold symmetry of the adsorption site and $A_{HF}$ = 3.7 μeV [12]. $\{B_m^n\}$ and $g_{eff}$ are the Steven's parameters and effective electron Landé $g$-factors, respectively, used in Ref [10]. $J$ and $I$ are the total angular momentum and nuclear spin operators, respectively, while $J_Z$ and $B_Z$ are the total spin and magnetic field, respectively, both projected along the out-of-plane direction. We do not include the nuclear contribution to the Zeeman term as the nuclear Landé g-factor is negligible. The states in the total spin level diagrams [Figure 3(a,d)] are labeled according to their legacy states.

**Acknowledgements**

P.R.F. acknowledges support from the Fulbright U.S. Student Program. F.D.N. greatly appreciates support from the Swiss National Science Foundation under project number PZ00P2_176866. P.R.F. thanks Michael Kopreski for useful discussion.

**Author information**

H.B. and F.D.N. supervised the work. F.D.N. conceived the project. P.R.F. and F.D.N. designed the experiments. F.P., D.P.S., and P.R.F. performed the measurements. P.R.F., E.F., D.P.S., and F.D.N. analyzed the data. P.R.F. and F.D.N. implemented the EasySpin model. P.R.F. developed the linear field sweep routines. P.R.F., F.D.N., and H.B. wrote the manuscript. All authors discussed the results and contributed to the manuscript.

**Competing interests**

The authors declare no competing interests.

**Figure Captions**

**Figure 1 | Measuring magnetic states of Ho with an antiferromagnetic tip. a,** Schematic of antiferromagnetic tip (green) tunneling into a holmium single atom magnet above (left and right) and below (middle) the switching threshold, $V_t$. These three phases are labelled 'read', 'evolve', and 'read'. Semitransparent schematic shows retracted tip used later for zero field studies. STM topographic image of Ho adsorbed on O top site of MgO/Ag(100) ($I = 104$ pA, $V = -130$ mV). Scale bar (white) is 2 nm. **b,** current-time trace showing two-state switching of Ho ($I_{set} = 100$ pA, $V = -130$ mV, Z-feedback open). The magnetic state (Up) from the left panel is retained after tunneling below the switching threshold for 16 minutes as seen in the right panel.

**Figure 2 | Zero magnetic field evolution experiment. a,** Schematic of tip height (upper), tip bias (middle), and tunneling current (lower) as a function of time for one retract/approach cycle. The purple and green regions highlight periods during which the magnetic state is randomized and simultaneously read. The yellow region highlights the period of zero field evolution for a duration of $\Delta t$. The semitransparent blue curve in the middle panel shows the multiple retract/approach cycles described in the text with the tip bias below the switching threshold, $V_t$. The three plots share the time axis. **b,** Representative current-time traces for the four observed outcomes: Up to Up (first from top), Down to Down (second), Up to Down (third), and Down to Up (fourth) ($I_{set} = 100$ pA, $V = -120$ mV (upper two traces), $V = 130$ mV (lower two traces), Z-feedback

open). Vertical scale bars correspond to 5 pA. **c**, Hold-time, $\Delta t$, dependent occurrence of magnetic state retention for the Up (green) and Down (red) states and magnetic state reversal for traces beginning in the Up (blue) and Down (orange) states at zero field. **d**, Plot of magnetic state retention/reversal as a function of the number, $N$, of retract/approach cycles. Same color coding as **c**.

**Figure 3 | Comparing the $J_Z$=8 and $J_Z$=7 models. a,** Extended Zeeman diagram for the $J_Z$=8 electronic ground state highlighting the 8 avoided level crossings (red circles) and the lowest energy zero field crossing (black square). Inset: Total spin level diagram with $J_Z$=8 (orange), $J_Z$=7 (blue), and other total spin legacy states (green). **b,** Lowest energy zero field crossing (region enclosed in black square in **a**) with (solid orange) and without (dashed red) nuclear spin for $J_Z$=8 model. **c,** Representative avoided level crossing (red circles in **a**) for $J_Z$=8 model. **d,** Same as **a** for $J_Z$=7 model with lowest energy zero field crossing (purple square) and highest magnetic field crossing (gold circle). Inset: same as inset in **d** for $J_Z$=7 case. **e,** Lowest energy zero field crossing (region enclosed in purple square in **d**) for $J_Z$=7 model. **f,** Real level crossing (red circle in **d**) for $J_Z$=7 model.

**Table 1 | State inversion measurement.** Observed occurrence of magnetic state reversal for the linear field sweep protocol described in the main text with the first column denoting the observed magnetic state before and after the retract/approach protocol and the first row describing the speeds at which the tip was retracted and then approached.

**Figure 1 |**

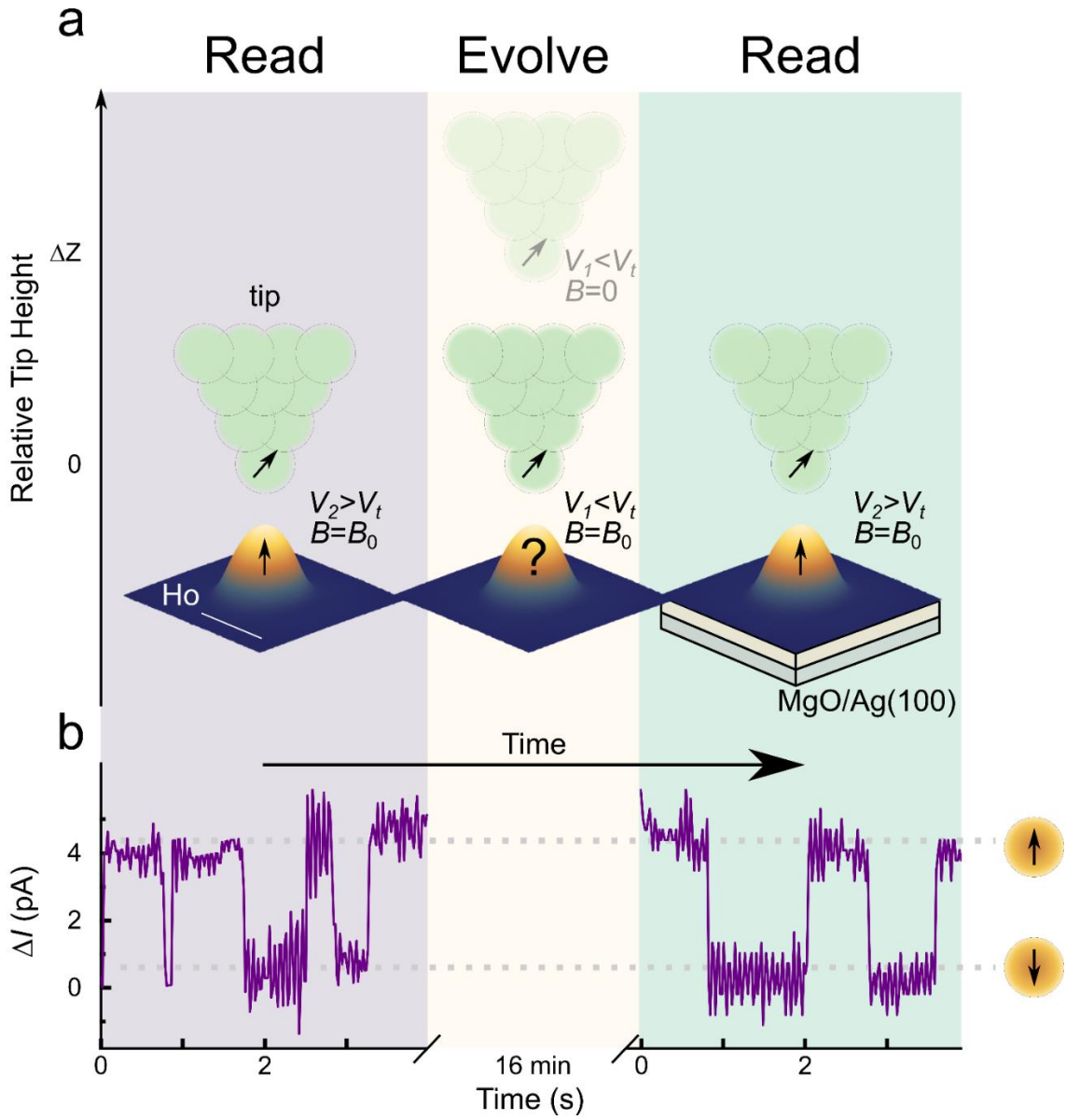

*Figure 1: Concept and Topo*

**Figure 2 |**

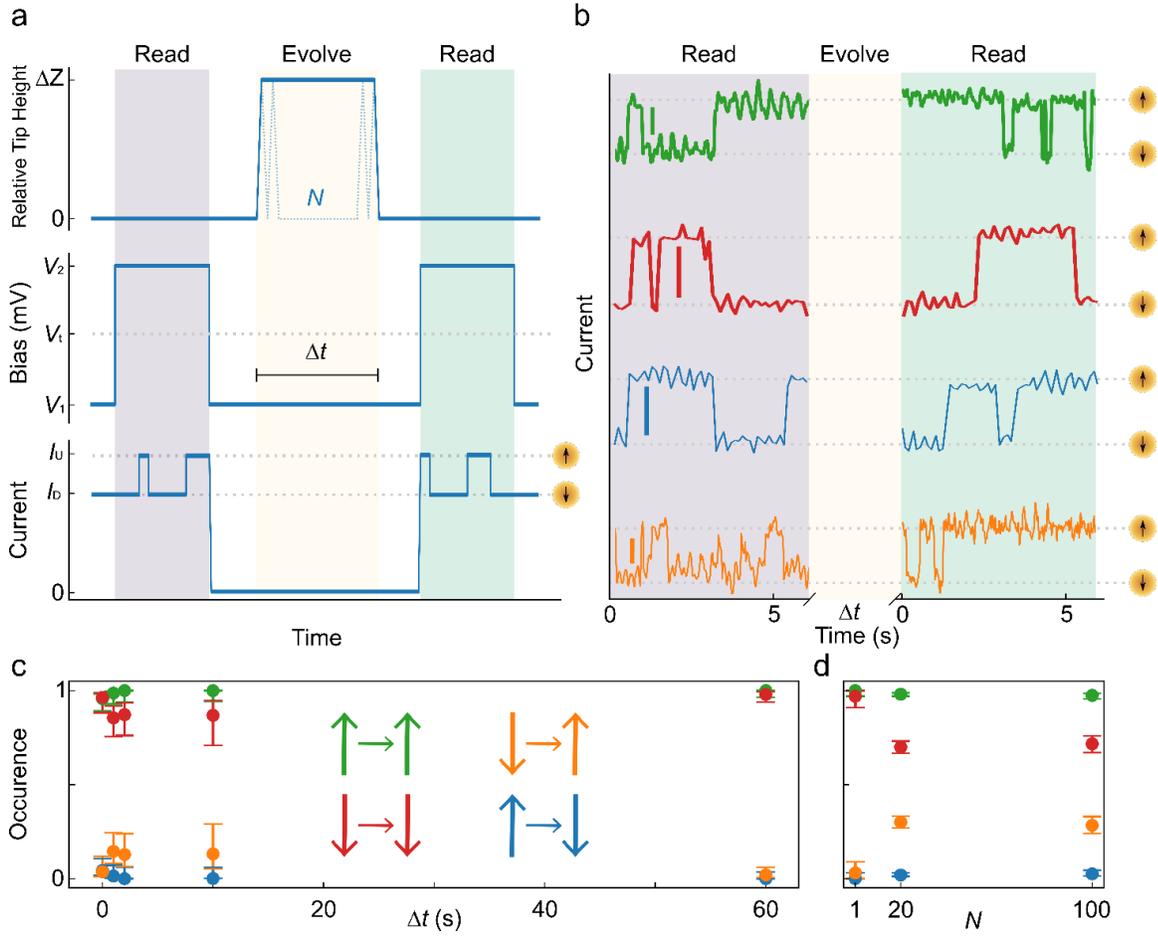

*Figure 2: Zero Field Stability*

**Figure 3 |**

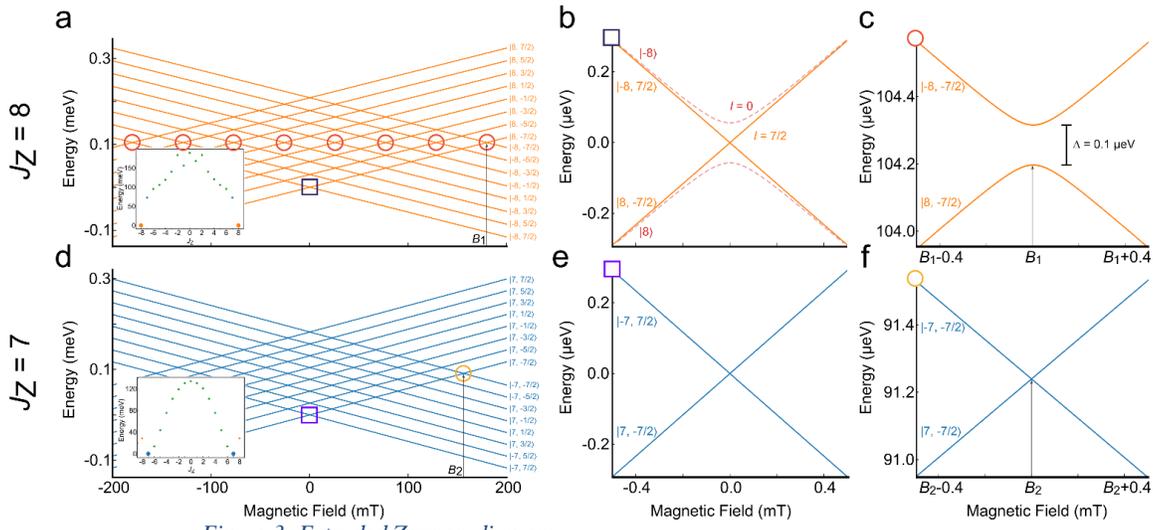

*Figure 3: Extended Zeeman diagram*

**Table 1 |**

|  | Slow-Fast | Fast-Slow |
|---|---|---|
| Up to down | $55^{+15}_{-16}\%$ | $35^{+19}_{-15}\%$ |
| Down to up | $48^{+17}_{-17}\%$ | $59^{+20}_{-23}\%$ |